\documentclass[10pt]{article}
\usepackage{amsmath,amssymb}
\usepackage{graphicx}
\usepackage{hyperref}
\DeclareGraphicsExtensions{.eps,.bmp,.wmf,.jpg,.pdf}
\numberwithin{equation}{section}
\def\be{\begin{equation}}
\def\ee{\end{equation}}

\def\bea{\begin{eqnarray}}
\def\eea{\end{eqnarray}}
\title{\textbf{A viable $f(R)$ gravity model without oscillations in the effective dark energy}}
\author{A. Oliveros\thanks{alexanderoliveros@mail.uniatlantico.edu.co}\\
Programa de F\'isica, Universidad del Atl\'antico, Carrera 30 N\'umero 8-49\\
Puerto Colombia-Atl\'antico, Colombia} 

\date{}
\begin{document}
\maketitle

\begin{abstract}
In this study, we propose a reparameterization of a specific viable $f(R)$ gravity model to represent it as a perturbation of the $\Lambda$CDM model. The $f(R)$ gravity model under consideration includes two parameters, $b$ and $n$, which control how close the proposed model can be to $\Lambda$CDM, allowing for arbitrary proximity. Furthermore, it is shown that the Hu-Sawicki (HS) model is a limiting case of this reparameterized model. Following the existing literature, we also derive an analytical approximation for the expansion rate $H(z)$, which shows an excellent agreement between this analytical approximation and the numerical solution over a wide range of redshifts for realistic values of the deviation parameter $b$.
By appropriately selecting values for the model parameters, we plot the cosmological parameters $w_{\rm{DE}}$, $w_{\rm{eff}}$, $\Omega_{\rm{DE}}$, and $H(z)$, as well as the statefinder quantities $q$, $j$, $s$, and $Om(z)$. We find that their present values (at $z=0$) are consistent with the observations from Planck 2018 and the values predicted by the $\Lambda$CDM model.
It is important to note that the examined cosmological and statefinder parameters do not exhibit significant oscillations of effective dark energy, which could lead to singular and unphysical solutions at high redshifts. This anomalous behavior has been avoided here by utilizing the approximate analytical solution for $H(z)$.
Additionally, we conduct a detailed analysis of the evolution of matter density perturbations within the introduced $f(R)$ gravity model. The results demonstrate that this viable $f(R)$ gravity model is practically indistinguishable from the $\Lambda$CDM model at the background level.
\end{abstract}


\section{Introduction}\label{sec_intro}
\noindent The discovery of late-time cosmic acceleration in the late 1990s has attracted a huge attention from the scientific community over the past few decades, and it has prompted extensive efforts to explain it. The commonly proposed explanation for this late-time cosmic acceleration is the introduction of a new energy component in the Universe known as dark energy (DE), characterized by a negative pressure. However, to date, there is no satisfactory solution to the DE problem, and its explanation within fundamental theories of physics remains elusive (for a comprehensive review about this topic see Refs.~\cite{peebles,copeland,bamba1}). Within the wide range of proposals attempting to provide an explanation for the DE problem, modified gravity theories have positioned themselves as an excellent alternative. In this context, it is not necessary to consider DE or new forms of matter to explain the late-time acceleration (see e.g.~Refs. \cite{odintsov1,clifton,
odintsov2} for a review). A modified gravity theory that has been successfully applied in the cosmological and astrophysical context is
the $f(R)$ theory of gravity. In this case, the late-time cosmic acceleration arises as an effect with geometrical origin
\cite{hwang, nojiri1, capozziello1, nojiri2, capozziello2, nojiri3, hu1, mao, olmo, hu2, faraoni, bean, nojiri4, capozziello3, sasaki, appleby, carloni, capozziello4, elizalde, capozziello5, odintsov3, odintsov4, odintsov5, oikonomou1, oikonomou2}. However, the selection of a specific $f(R)$ function is not arbitrary; it must adhere to several consistency requirements and various constraints that impose conditions for the cosmological viability of $f(R)$ dark energy models \cite{amendola}. Despite the success of $f(R)$ gravity, certain difficulties have arisen during the analysis of some specific models. For example, some $f(R)$ gravity models exhibit large frequency oscillations of effective dark energy, which makes solutions singular and unphysical at a high redshift. Addressing this issue,  Ref.~\cite{elizalde1} analyzes the oscillatory behavior of the $f(R)$ dark energy around the phantom divide line, $w_{\rm{DE}}=-1$, both during the matter era and also in the de Sitter epoch. Similarly,  Ref.~\cite{bamba} investigates this characteristic in viable $f(R)$ gravity models, specifically exploring exponential gravity and a power form model. In this case, the authors propose improving the models by introducing a corrective term that effectively stabilizes the oscillations without compromising the viability properties. In the same way,  Ref.~\cite{odintsov7} examines a similar behavior within an $f(R)$ Einstein-Gauss-Bonnet gravity model, while Ref.~\cite{fronimos} considers an  $f(R)$ Einstein-Gauss-Bonnet model with a non-minimal coupling between gravity and the kinetic term of a scalar field.  Lastly, Ref.~\cite{oikonomou3} delves into the the late-time dynamics of several $f(R)$ gravity models that exhibit this peculiar behavior.\\

\noindent In the present study, we aim to demonstrate that it is possible to circumvent this issue within a specific viable $f(R)$ gravity model. Building upon existing literature, we employ a reparameterization technique to express this model as a perturbation from the $\Lambda$CDM model. The $f(R)$ gravity model under consideration includes two parameters, $b$ and $n$, and it is demonstrated that the closeness of this model to $\Lambda$CDM depends on these parameters, allowing for arbitrary proximity. Subsequently, we perform an analytical perturbative expansion in the parameter $b$, yielding an analytic expression for $H(\Omega_{m0},b;z)$ to all orders in $b$ (for further details, refer to Ref.~\cite{basilakos}). This approach has also been used in other contexts where an analytical expression for $H(\Omega_{m0},b;z)$ allows for easier modifications of computational codes, like CLASS. (see Refs. \cite{nesseris1} and \cite{nesseris2}
for more details). The $f(R)$ gravity model being reparameterized was originally introduced in Ref.~\cite{granda}, and its late-time cosmological evolution has been extensively studied in Ref.~\cite{oliveros}.
Furthermore, by employing the analytic expression for $H(\Omega_{m0},b;z)$ and appropriately selecting values for the model parameters, we plot various cosmological parameters, including $w_{\rm{DE}}$, $w_{\rm{eff}}$, $\Omega_{\rm{DE}}$, and $H(z)$, alongside statefinder quantities such as $q$, $j$, $s$, and $Om(z)$. Moreover, we compare their present values (at $z=0$) with observations from Planck 2018 and the predicted values of the $\Lambda$CDM model. The primary goal is to demonstrate that the examined cosmological and statefinder parameters do not exhibit significant oscillations of effective dark energy, which could result in singular and unphysical solutions at high redshifts. Finally, we conduct a comprehensive analysis of the evolution of matter density perturbations within the considered $f(R)$ gravity model.\\

\noindent This paper is organized as follows: In Section \ref{model}, we provide a concise overview of $f(R)$ gravity. Additionally, we carry out the reparameterization of a particular $f(R)$ gravity model and conduct an analytical perturbative expansion in the parameter $b$. This expansion leads to an analytic expression for $H(\Omega_{m0},b;z)$ at all orders in $b$. In Section \ref{cosmo-analysis}, we plot the cosmological parameters, namely $w_{\rm{DE}}$, $w_{\rm{eff}}$, $\Omega_{\rm{DE}}$, and $H(z)$, as well as the statefinder quantities $q$, $j$, $s$, and $Om(z)$, while considering selected values for the model parameters. We calculate their present values at $z=0$ and compare them with observations from Planck 2018, as well as the values predicted by the $\Lambda$CDM model. Furthermore, we provide a comprehensive analysis of the evolution of matter density perturbations within the $f(R)$ gravity model introduced. Our conclusions are presented in Section \ref{conclus}.
\section{The model}\label{model} 
\noindent In general, the action for an $f(R)$ gravity model in the presence  of matter components is given by
\begin{equation}\label{eq1}
S=\int{d^4x\sqrt{-g}\left(\frac{f(R)}{2\kappa^2}+\mathcal{L}_M\right)},
\end{equation}
where $g$ denotes the determinant of the metric tensor $g^{\mu\nu}$, $\kappa^2=8\pi G=1/M_p^2$, with $G$ being the Newton's constant and $M_p$ the reduced Planck mass. $\mathcal{L}_M$ represents the Lagrangian density for the matter components (relativistic and non-relativistic perfect matter
fluids). The term $f(R)$ is for now an arbitrary function of the Ricci scalar $R$. Variation with respect to the metric gives the equation
of motion
\begin{equation}\label{eq2}
f_R(R)R_{\mu\nu}-\frac{1}{2}g_{\mu\nu}f(R)+(g_{\mu\nu}\square-\nabla_\mu\nabla_\nu)f_R(R)=\kappa^2T_{\mu\nu}^{(M)},
\end{equation}
where $f_R\equiv \frac{df}{dR}$, $\nabla_{\mu}$ is the covariant derivative associated with the Levi-Civita connection of the metric, and
$\square\equiv \nabla^\mu\nabla_\mu$. Plus, $T_{\mu\nu}^{(M)}$ is the matter energy–momentum tensor which is assumed to be a perfect fluid.
Considering the flat Friedman-Robertson-Walker (FRW) metric,
\begin{equation}\label{eq3}
ds^2=-dt^2+a(t)^2\delta_{ij}dx^idx^j,
\end{equation}
with $a(t)$ representing the scale factor,  the time and spatial components of Eq.~(\ref{eq2}) are given, respectively, by
\begin{equation}\label{eq4}
3H^2f_R=\kappa^2(\rho_{\rm{m}}+\rho_{\rm{r}})+\frac{1}{2}(Rf_R-f)-3H\dot{f}_R,
\end{equation}
and
\begin{equation}\label{eq5}
-2\dot{H}f_R=\kappa^2\left(\rho_{\rm{m}}+\frac{4}{3}\rho_{\rm{r}}\right)+\ddot{f}_R-H\dot{f}_R,
\end{equation}
where $\rho_{\rm{m}}$ is the matter density and $\rho_{\rm{r}}$ denotes the density of radiation. The over-dot denotes a derivative with respect to the cosmic time $t$ and $H\equiv \dot{a}/a$ is the Hubble parameter. If there is no interaction between non-relativistic
matter and radiation, then these components obey separately the conservation laws
\begin{equation}\label{eq6}
\dot{\rho}_{\rm{m}}+3H\rho_{\rm{m}}=0,\quad \dot{\rho}_{\rm{r}}+4H\rho_{\rm{r}}=0.
\end{equation}
As usual in the literature, it is possible to rewrite the field equations (\ref{eq4}) and (\ref{eq5}) in the Einstein-Hilbert form:
\begin{equation}\label{eq6'}
3H^2=\kappa^2\rho,
\end{equation}
\begin{equation}\label{eq7'}
-2\dot{H}^2=\kappa^2(\rho+p),
\end{equation}
where $\rho=\rho_{\rm{m}}+\rho_{\rm{r}}+\rho_{\rm{DE}}$ and $p=p_{\rm{m}}+p_{\rm{r}}+p_{\rm{DE}}$ correspond to the total effective energy density and total effective pressure density of the cosmological fluid.  In this case, the dark energy component has a geometric origin, and after a some manipulation in Eqs. (\ref{eq4})  and (\ref{eq5}), we obtain the effective dark energy and pressure corresponding to $f(R)$-theory given by
\begin{equation}\label{eq8'}
\rho_{\rm{DE}}=\frac{1}{\kappa^2}\left[\frac{Rf_R-f}{2}+3H^2(1-f_R)-3H\dot{f}_R\right],
\end{equation}
and
\begin{equation}\label{eq9'}
p_{\rm{DE}}=\frac{1}{\kappa^2}[\ddot{f}_R-H\dot{f}_R+2\dot{H}(f_R-1)-\kappa^2\rho_{\rm{DE}}],
\end{equation}
it is easy to show that $\rho_{\rm{DE}}$ and $p_{\rm{DE}}$ defined in this way satisfy the usual energy conservation equation
\begin{equation}\label{eq10'}
\dot{\rho}_{\rm{DE}}+3H(\rho_{\rm{DE}}+p_{\rm{DE}})=0,
\end{equation}
in this case, we assume that the equation of state parameter for this effective dark energy satisfies the following relation:
\begin{equation}\label{eq11'}
w_{\rm{DE}}=\frac{p_{\rm{DE}}}{\rho_{\rm{DE}}},
\end{equation}
now, it is well known that the Ricci scalar can be expressed in terms of the Hubble parameter as
\begin{equation}\label{eq7}
R=6(2H^2+\dot{H}).
\end{equation}
The $f(R)$ gravity model, which plays a central role in this work, is derived from the following $f(R)$ model:
\begin{equation}\label{eq8}
f(R)=R-2\,\lambda\,\mu^2\,e^{-(\mu^2/R)^{n}},
\end{equation}
where $\lambda$ and $n$ are positive real dimensionless parameters, and $\mu$ is a positive real parameter with dimension of $\rm{eV}$.
This model was introduced in Ref. \cite{granda}, and it behaves very close to $\Lambda$CDM at early times and satisfy local and cosmological constraints.  In Ref. \cite{oliveros} the authors analyze the late-time evolution of the Universe for this model and perform an statistical analysis to constrain the free parameters of the model.
An extension of this model (where an $R^2$ Starobinsky term is added) able to explain early time inflation and late time accelerated expansion was studied in Ref.~\cite{granda2}. Although such an extension, i.e., including an extra $R^2$ term in Eq. (\ref{eq8}), might be of some interest (see \cite{odintsov10} for a recent discussion about this subject), this term does not have relevant contribution at late-times (dark energy era). Then, in  Ref. \cite{granda3}, the author performs a generalization of this model, introducing a general function of the scalar curvature in the exponential term. In the literature, other authors have studied some $f(R)$ gravity models with exponential functions of the scalar curvature (see for example Refs. \cite{linder, odintsov8, Odintsov:2017qif, Odintsov:2018qug}).\\

\noindent Following the ideas worked out in Ref. \cite{basilakos}, in this work we reparameterize the model (\ref{eq8}) in order to express it as a perturbation deviating from the $\Lambda$CDM Lagrangian. In this sense, choosing $\mu^2=b\Lambda$ and $\lambda=1/b$, the Eq. (\ref{eq8})  reduce to
\begin{equation}\label{eq9}
f(R)=R-2\,\Lambda\,e^{-(b\Lambda/R)^n},
\end{equation}
and in this form it is evident that  this model can be arbitrarily close to $\Lambda$CDM, depending on the parameters $b$ and $n$.
This statement clearly illustrates that the reason for successfully passing all the observational tests  is primarily due to small perturbations around the $\Lambda$CDM model. Furthermore
\begin{equation}\label{eq10}
\lim_{b\rightarrow 0}f(R)=R-2\Lambda,
\end{equation}
\begin{equation}\label{eq11}
\lim_{b\rightarrow \infty}f(R)=R,
\end{equation}
From this, we can say that this model converges to $\Lambda$CDM for $b\rightarrow 0$, while for $b\rightarrow \infty$, the model give rise to a matter dominated universe. Therefore, it is obvious that this model contain the cosmological constant $\Lambda$ and in this way, the model must satisfy the solar system tests. In this sense, in Ref. \cite{nesseris} the authors carried out cosmological constraints on several well-known $f(R)$ models, but also on a new class of models that are variants of the Hu-Sawicki model (HS) \cite{hu} one of the form
\begin{equation}\label{eq11'}
f(R)=R-\frac{2\Lambda}{1+b\,y(R,\Lambda)},
\end{equation}
which interpolates between the cosmological constant model and a matter dominated universe for different values of the
parameter $b$. It is worth saying that Eq. (\ref{eq11'}) represents a specific case of the more general form of the $f(R)$ function \cite{basilakos, nesseris, bamba}:
\begin{equation}\label{eq12'}
f(R)=R-2\Lambda\tilde{y}(R,b);
\end{equation}
is evident that our model given by Eq. (\ref{eq9}) has the same form as the above.\\

\noindent It is interesting to see that the $f(R)$ model given by Eq. (\ref{eq9}) contains the HS model as a 
limiting case. In Ref. \cite{basilakos}, the authors simplify the HS model through simple algebraic manipulations and they get
\begin{equation}\label{eq12}
f(R)=R-\frac{2\Lambda}{1+\left(\frac{b\Lambda}{R}\right)^n},
\end{equation}
which can be rewritten as
\begin{equation}\label{eq13}
f(R)=R-2\Lambda\left[1+\left(\frac{b\Lambda}{R}\right)^n\right]^{-1},
\end{equation}
and assuming that $(b\Lambda/R)^n\ll 1$, then the HS model reduces to
\begin{equation}\label{eq14}
f(R)=R-2\Lambda\left[1-\left(\frac{b\Lambda}{R}\right)^n\right],
\end{equation}
which corresponds to the result obtained by expanding Eq. (\ref{eq9}) in a Taylor expansion up to the first-order approximation.\\

\noindent Another significant advantage of the model given by Eq. (\ref{eq9}) is that we can obtain an analytic approximation for the expansion rate $H(z)$. Following the procedure carried out in Ref. \cite{basilakos}, we rewrite Eq. (\ref{eq4}) in terms of
$N=\ln{a}$
\begin{equation}\label{eq15}
-f_RH^2(N)+(\Omega_{m0}e^{-3N}+\Omega_{r0}e^{-4N})H_0^2+\frac{1}{6}(Rf_R-f)=f_{RR}H^2(N)R'(N),
\end{equation}
here prime denotes differentiation with respect to $N$. Additionally, Eq. ~(\ref{eq7}) is given by
\begin{equation}\label{eq16}
R(N)=6\left[2H^2(N)+\frac{1}{2}\frac{dH^{2}(N)}{dN}\right],
\end{equation}
since the model under study here approach to $\Lambda$CDM as $b\rightarrow 0$, one can express the solution to Eq. (\ref{eq15}), $H^2(N)$, as a Taylor expansion in the deviation parameter, $b$, as follows
\begin{equation}\label{eq17}
H^2(N)=H^2_{\Lambda}(N)+\sum_{i=1}^{M}b^i\delta H_i^2(N),
\end{equation}
where
\begin{equation}\label{eq18}
\frac{H^2_{\Lambda}(N)}{H_0^2}=\Omega_{m0}\,e^{-3N}+\Omega_{r0}\,e^{-4N}+(1-\Omega_{m0}-\Omega_{r0})=E_{\Lambda}^2(N),
\end{equation}
further, as has been demonstrate in Ref.\cite{basilakos}, we can consider only two terms in the above series expansion, in this way 
Eq. (\ref{eq17}) reduces to
\begin{equation}\label{eq19}
H^2(N)\approx H^2_{\Lambda}(N)+b\,\delta H_1^2(N)+b^2\delta H_2^2(N),
\end{equation}
and considering by simplicity $n=1$ in our model, then, $\delta H_1^2(N)$ and $\delta H_2^2(N)$, are given by
\begin{equation}\label{eq20}
\frac{\delta H_1^2(N)}{H_0^2}=\frac{-2H_0^2(1-\Omega_{m0}-\Omega_{r0})^2\left[12(H_{\Lambda}^2(N))^2+\left(\frac{dH_{\Lambda}^{2}(N)}{dN}\right)^2+H^2_{\Lambda}(N)\left(15\frac{dH_{\Lambda}^{2}(N)}{dN}+2\frac{d^2H_{\Lambda}^{2}(N)}{dN^2}\right)\right]}{\left[4H^2_{\Lambda}(N)+\frac{dH_{\Lambda}^{2}(N)}{dN}\right]^3},
\end{equation}
\begin{equation}\label{eq21}
\begin{aligned}
\frac{\delta H_2^2(N)}{H_0^2}=&\bigg[H_0^2(1-\Omega_{m0}-\Omega_{r0})^3\bigg(8192(H_{\Lambda}^2(N))^6+3\left(\frac{dH_{\Lambda}^{2}(N)}{dN}\right)^6+8H_0^2(1-\Omega_{m0}-\Omega_{r0})\left(\frac{dH_{\Lambda}^{2}(N)}{dN}\right)^4\\
&\times\frac{d^2H_{\Lambda}^{2}(N)}{dN^2}+1024(H_{\Lambda}^2(N))^5\bigg(25\frac{dH_{\Lambda}^{2}(N)}{dN}+3\bigg(\frac{d^2H_{\Lambda}^{2}(N)}{dN^2}-6(1-\Omega_{m0}-\Omega_{r0})\bigg)\bigg)\\
&+4H_{\Lambda}^2(N)\left(\frac{dH_{\Lambda}^{2}(N)}{dN}\right)^2\bigg(29\left(\frac{dH_{\Lambda}^{2}(N)}{dN}\right)^3-24(1-\Omega_{m0}-\Omega_{r0})\left(\frac{d^2H_{\Lambda}^{2}(N)}{dN^2}\right)^2\\
&+3\left(\frac{dH_{\Lambda}^{2}(N)}{dN}\right)^2\left(\frac{d^2H_{\Lambda}^{2}(N)}{dN^2}-198(1-\Omega_{m0}-\Omega_{r0})\right)-8H_0^2(1-\Omega_{m0}-\Omega_{r0})\frac{dH_{\Lambda}^{2}(N)}{dN}\\
&\times\left(22\frac{d^2H_{\Lambda}^{2}(N)}{dN^2}-\frac{d^3H_{\Lambda}^{2}(N)}{dN^3}\right)\bigg)+256(H_{\Lambda}^2(N))^4\bigg(83\left(\frac{dH_{\Lambda}^{2}(N)}{dN}\right)^2+12\frac{dH_{\Lambda}^{2}(N)}{dN}\\
&\times\left(\frac{d^2H_{\Lambda}^{2}(N)}{dN^2}-24H_0^2(1-\Omega_{m0}-\Omega_{r0})\right)-2H_0^2(1-\Omega_{m0}-\Omega_{r0})\bigg(15\frac{d^2H_{\Lambda}^{2}(N)}{dN^2}-6\frac{d^3H_{\Lambda}^{2}(N)}{dN^3}\\
&-\frac{d^4H_{\Lambda}^{2}(N)}{dN^4}\bigg)\bigg)+128(H_{\Lambda}^2(N))^3\bigg(61\left(\frac{dH_{\Lambda}^{2}(N)}{dN}\right)^3+\left(\frac{dH_{\Lambda}^{2}(N)}{dN}\right)^2\bigg(9\frac{d^2H_{\Lambda}^{2}(N)}{dN^2}\\
&-510H_0^2(1-\Omega_{m0}-\Omega_{r0})\bigg)-3H_0^2(1-\Omega_{m0}-\Omega_{r0})\frac{d^2H_{\Lambda}^{2}(N)}{dN^2}\bigg(19\frac{d^2H_{\Lambda}^{2}(N)}{dN^2}+4\frac{d^3H_{\Lambda}^{2}(N)}{dN^3}\bigg)\\
&-2H_0^2(1-\Omega_{m0}-\Omega_{r0})\frac{dH_{\Lambda}^{2}(N)}{dN}\bigg(147\frac{d^2H_{\Lambda}^{2}(N)}{dN^2}+16\frac{d^3H_{\Lambda}^{2}(N)}{dN^3}-\frac{d^4H_{\Lambda}^{2}(N)}{dN^4}\bigg)\bigg)\\
&-32(H_{\Lambda}^2(N))^2\bigg(-44\left(\frac{dH_{\Lambda}^{2}(N)}{dN}\right)^4+\left(\frac{dH_{\Lambda}^{2}(N)}{dN}\right)^3\bigg(-672H_0^2(1-\Omega_{m0}-\Omega_{r0})\\
&-6\frac{d^2H_{\Lambda}^{2}(N)}{dN^2}\bigg)-21H_0^2(1-\Omega_{m0}-\Omega_{r0})\left(\frac{d^2H_{\Lambda}^{2}(N)}{dN^2}\right)^3-3H_0^2(1-\Omega_{m0}-\Omega_{r0})\frac{dH_{\Lambda}^{2}(N)}{dN}\\
&\times\frac{d^2H_{\Lambda}^{2}(N)}{dN^2}\left(61\frac{d^2H_{\Lambda}^{2}(N)}{dN^2}-4\frac{d^3H_{\Lambda}^{2}(N)}{dN^3}\right)-H_0^2(1-\Omega_{m0}-\Omega_{r0})\left(\frac{dH_{\Lambda}^{2}(N)}{dN}\right)^2\\
&\times\left(637\frac{d^2H_{\Lambda}^{2}(N)}{dN^2}-34\frac{d^3H_{\Lambda}^{2}(N)}{dN^3}+\frac{d^4H_{\Lambda}^{2}(N)}{dN^4}\right)\bigg)\bigg)\bigg]/\left (2\left(4H_{\Lambda}^2(N)+\frac{dH_{\Lambda}^{2}(N)}{dN}\right)^8\right),
\end{aligned}
\end{equation}
where, in order to simplify the calculations, we have expressed Eqs. (\ref{eq20})  and (\ref{eq21})  in terms of $H_{\Lambda}^{2}(N)$ and its derivatives. It is straightforward to show that Eq. (\ref{eq20}) is equivalent to the one obtained in Ref. \cite{basilakos} for the HS model. This similarity is not surprising, as demonstrated earlier, our model closely resembles the HS model at first order in $b$. Regarding Eq. (\ref{eq21}), it exhibits the same dependence on $H_{\Lambda}^{2}(N)$ and its derivatives as the equation derived in Ref. \cite{basilakos}. However, in our case, there are certain numerical coefficients that differ.\\

\begin{figure*}
\centering
    \includegraphics[width=0.47\textwidth]{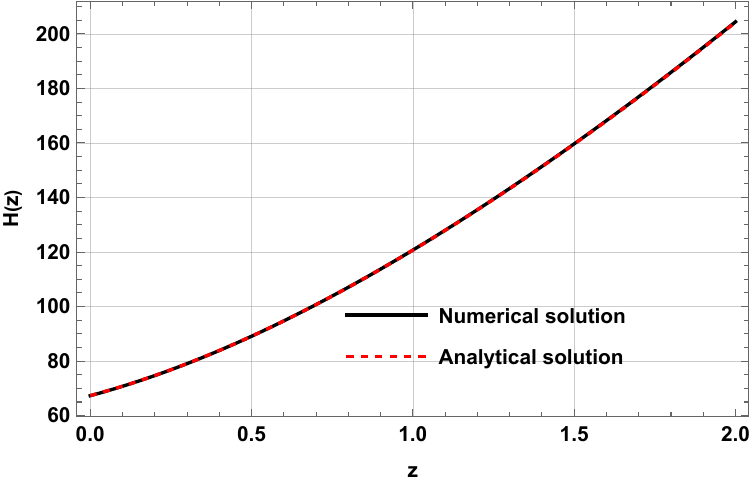}
    \includegraphics[width=0.47\textwidth]{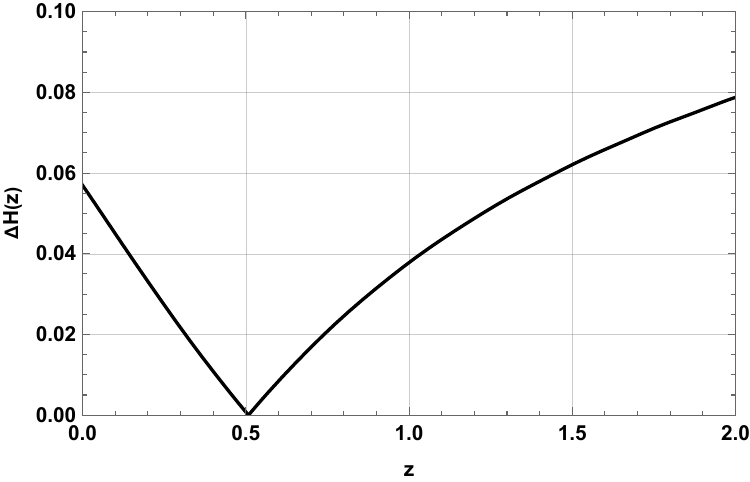}
    \caption{(\emph{left}) Plot for $H(z)$ vs. $z$ taking into account the numerical and analytical solutions (in units of $\rm{Km}\,\rm{s}^{-1}\,\rm{Mpc}^{-1}$). (\emph{right}) Comparison between the approximate analytical solution given by Eq. (\ref{eq22}) and the numerical solution for $H(z)$ obtained using the formalism described in  Ref. \cite{oliveros}. The values for the model parameters used here are: 
$\Omega_{m0}=0.3147$, $H_0=67.4\,\rm{Km}\,\rm{s}^{-1}\,\rm{Mpc}^{-1}$ and $b=0.01$.}
    \label{fig1}
\end{figure*}

\noindent Replacing Eqs. (\ref{eq20})  and (\ref{eq21}) in Eq. (\ref{eq19}),  and using Eq. (\ref{eq18}), we obtain an approximate solution for the Hubble parameter $H(z)$:
\begin{equation}\label{eq22}
\begin{aligned}
E^2(z)=&\frac{H^2(z)}{H_0^2}=1-\Omega_{m0}+(1+z)^3\Omega_{m0}\\
&+\frac{6b(\Omega_{m0}-1)^2\left(-4+\Omega_{m0}(9-3\Omega_{m0}+z(3+z(z+3))
(1+(3+2z(3+z(z+3)))\Omega_{m0}))\right)}{(4+(-3+z(3+z(z+3)))\Omega_{m0})^3}\\
&+\frac{b^2(\Omega_{m0}-1)^3}{(1+z)^{24}\left(\frac{4(1-\Omega_{m0})}{(1+z)^3}+\Omega_{m0}\right)^8}
\bigg[5120(\Omega_{m0}-1)^6+9216(1+z)^3(\Omega_{m0}-1)^5\Omega_{m0}\\
&-30144(1+z)^6(\Omega_{m0}-1)^4\Omega_{m0}^2+31424(1+z)^9(\Omega_{m0}-1)^3\Omega_{m0}^3-9468(1+z)^{12}\\
&\times (\Omega_{m0}-1)^2\Omega_{m0}^4-4344(1+z)^{15}(\Omega_{m0}-1)\Omega_{m0}^5+\frac{37}{2}(1+z)^{18}\Omega_{m0}^6\bigg],
\end{aligned}
\end{equation}
for simplicity, we have assumed $\Omega_{r0}=0$ and made the substitution $N=-\ln{(1+z)}$. In Ref. \cite{sultana}, the authors derived a comparable expression for the Hubble rate $H(z)$ in the HS model. However, in our study, we have found that the numerical coefficients differ from those obtained in their work. In the left panel of Fig. \ref{fig1} we can see the evolution of $H(z)$ vs. $z$ taking into account the approximate analytical solution given by Eq. (\ref{eq22}) and the numerical solution for $H(z)$ obtained using the formalism described in  Ref. \cite{oliveros}. In the right panel of Fig. 1, the comparison between them is made. In this case, the comparison is performed using
the quantity  $\Delta H(z)$, which is defined by
\begin{equation}\label{eq23}
\Delta H(z)=\left[\frac{|H_A(z)-H_N(z)|}{H_N(z)}\right]\times 100,
\end{equation}
where $\rm{A}=\rm{Analytical}$ and $\rm{N}=\rm{Numerical}$. From above, we can say that the difference between these solutions
is $\Delta H(z)\sim 0.01\,\%$, which means that the approximation behaves very well in this scenario.
\section{Cosmological analysis in late-time}\label{cosmo-analysis}
\begin{figure*}
\centering
    \includegraphics[width=0.47\textwidth]{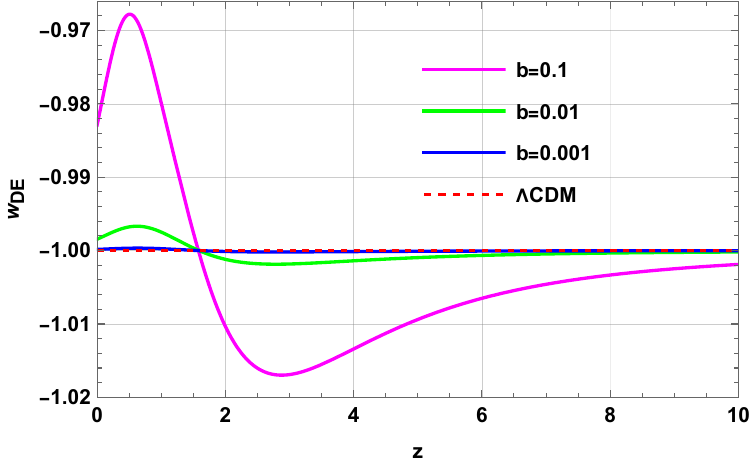}
    \includegraphics[width=0.47\textwidth]{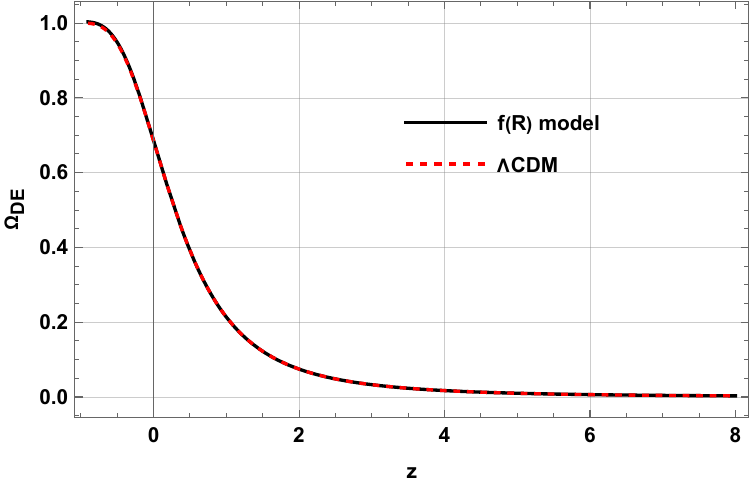}
    \caption{(\emph{left}) Evolution of $w_{\rm{DE}}$ vs. $z$ and (\emph{right}) $\Omega_{\rm{DE}}$ vs. $z$.}
    \label{fig2}
\end{figure*}

\noindent In this section, we will examine the aforementioned results in order to conduct a thorough analysis about the feasibility of the model presented in Eq. (\ref{eq9}). Additionally, we demonstrate the elimination of oscillations in the effective dark energy within this particular scenario. In this regard, we initiate the analysis by examining the behaviour of the cosmological parameters, namely
$w_{\rm{DE}}$, $w_{\rm{eff}}$, and $\Omega_{DE}$, as well as the statefinder quantities $q$, $j$, $s$, and $Om(z)$ in the late-time regime.  Let's remember that the cosmological parameters, $w_{\rm{DE}}$, $w_{\rm{eff}}$, and $\Omega_{\rm{DE}}$ in terms of the redshift $z$, are given by:
\begin{equation}\label{eq24}
w_{\rm{DE}}=-1+\frac{1}{3}(1+z)\frac{(\rho_{\rm{DE}}(z))'}{\rho_{\rm{DE}}(z)},
\end{equation}
\begin{equation}\label{eq25}
w_{\rm{eff}}=-1+\frac{1}{3}(1+z)\frac{(E^2(z))'}{E^2(z)},
\end{equation}
\begin{equation}\label{eq26}
\Omega_{\rm{DE}}=\frac{1-\Omega_{m0}}{E^2(z)};
\end{equation}
similarly,  the statefinder quantities $q$, $j$, $s$ and $Om(z)$ in terms of the redshift $z$, are given by:            
\begin{equation}\label{eq27}
q=-1+\frac{1}{2}(1+z)\frac{(E^2(z))'}{E^2(z)},
\end{equation}																
\begin{equation}\label{eq28}
j=-2-3q+\frac{(1+z)^2(E^2(z))''+(1+z)(E^2(z))'}{2E^2(z)},
\end{equation}																	
\begin{equation}\label{eq29}
s=\frac{j-1}{3\left(q-\frac{1}{2}\right)},
\end{equation}
\begin{equation}\label{eq30}
Om(z)=\frac{\frac{E^2(z)}{E^2(0)}-1}{(1+z)^3-1},
\end{equation}																
where prime denotes differentiation with respect to $z$. Using Eqs. (\ref{eq8'}) and (\ref{eq22}),	we can plot the above expressions in terms of the redshift $z$. Additionally,  in order to compare the results with the $\Lambda$CDM  model, we have also incorporated in these plots the corresponding behavior associated with each quantity defined by Eqs. (\ref{eq24})-(\ref{eq30}), but using  Eq. (\ref{eq18})  instead of (\ref{eq22}). In the left panel of Fig. \ref{fig2} we show the cosmological evolution of $w_{\rm{DE}}$ as a function of $z$ using some fixed values of $b$ and we can see that for large values of $z$, $w_{\rm{DE}}$ is very close to the $\Lambda$CDM model prediction (i.e. $w_{\Lambda}=-1$). For small values of $z$,  $w_{\rm{DE}}$ present a variation (jump) about -1, and we can see that as $b$ decreases,
$w_{\rm{DE}}$  progressively converges towards the $\Lambda$CDM model, i.e. this jump represents the deviation from $\Lambda$CDM model and it is directly related to the parameter $b$. A similar behavior was found for the HS model in Ref. cite{hu}. The equation of state crosses $w_{\rm{DE}}=-1$ at approximately the redshift $z\approx 1.59$. At early times, we have $1+w_{\rm{DE}}<0$ thus violating the strong energy conditions (SEC). A similar behavior was found for the HS model in Ref. \cite{nesseris1}. In the right panel of Fig. \ref{fig2}, we depict the evolution of $\Omega_{\rm{DE}}$ in terms of redshift $z$. From this, it is clear that for high redshifts, $\Omega_{\rm{DE}}$ lays to zero (matter dominated era) and for the distant future ($z\rightarrow -1$), $\Omega_{\rm{DE}}\rightarrow 1$ (de Sitter phase). 
\begin{figure*}
\centering
    \includegraphics[width=0.47\textwidth]{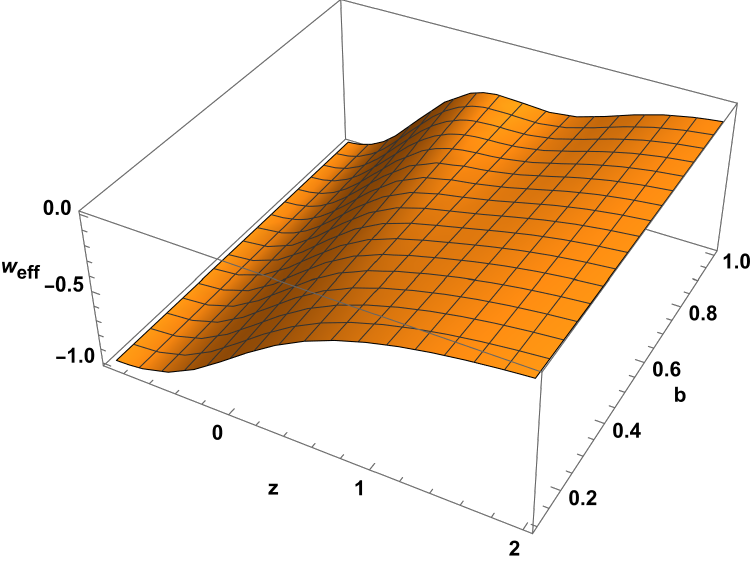}
    \includegraphics[width=0.47\textwidth]{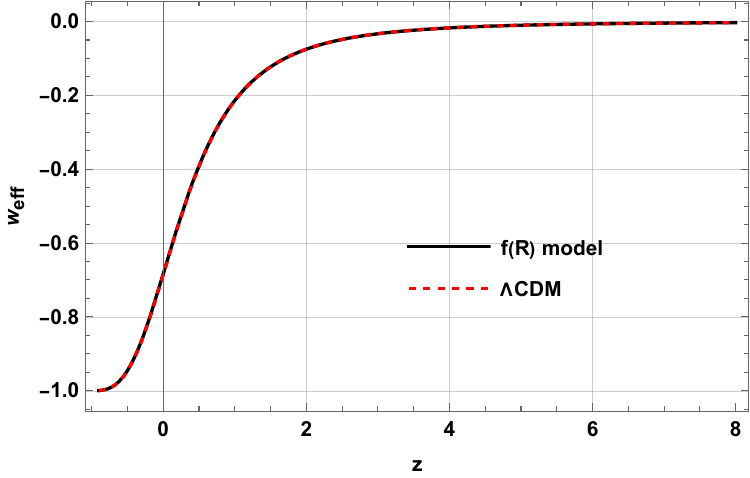}
    \caption{(\emph{left}) Evolution of $w_{\rm{eff}}$ as a function of $z$ and $b$. (\emph{right}) $w_{\rm{eff}}$ vs. $z$.}
    \label{fig3}
\end{figure*}	
Now, we depict the cosmological evolution of $w_{\rm{eff}}$ as a function of the redshift $z$ and the $b$ parameter in 
Fig. \ref{fig3}. 	We can see that, independently on the choice of $b$, $w_{\rm{eff}}$ starts from zero in the matter dominated era and asymptotically approaches -1 without any appreciable deviation. In the right panel of
Fig. \ref{fig3}, we represent 	$w_{\rm{eff}}$ in terms of $z$	considering a fixed value for $b$, in this case $b=0.01$, and is clear that	 in this scenario, we do not have the crossing of the phantom divide, which has been found for other $f(R)$ models due to the oscillatory behaviour of dark energy \cite{bamba}. In the  Figs. \ref{fig4}	and \ref{fig5}, the statefinder parameters	$q$, $j$, $s$, and $Om(z)$
present a behaviour very close to that obtained using the $\Lambda$CDM model. Therefore, the above results demonstrate that this viable $f(R)$ gravity model is practically indistinguishable from the $\Lambda$CDM model at the background level.\\
\begin{figure*}
\centering
    \includegraphics[width=0.47\textwidth]{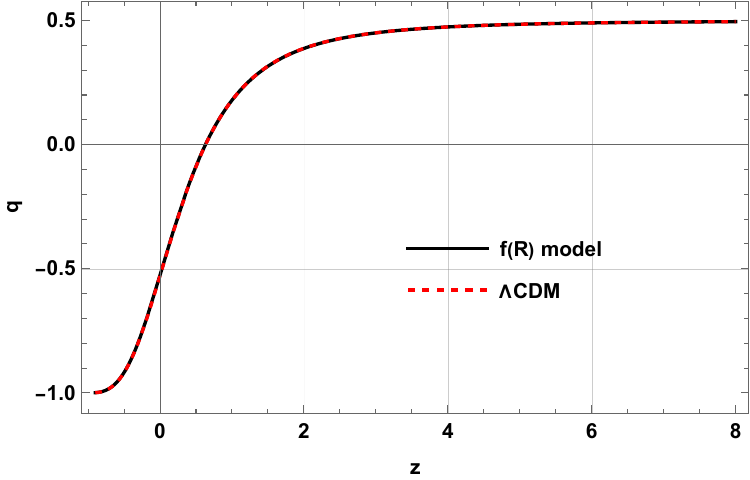}
    \includegraphics[width=0.47\textwidth]{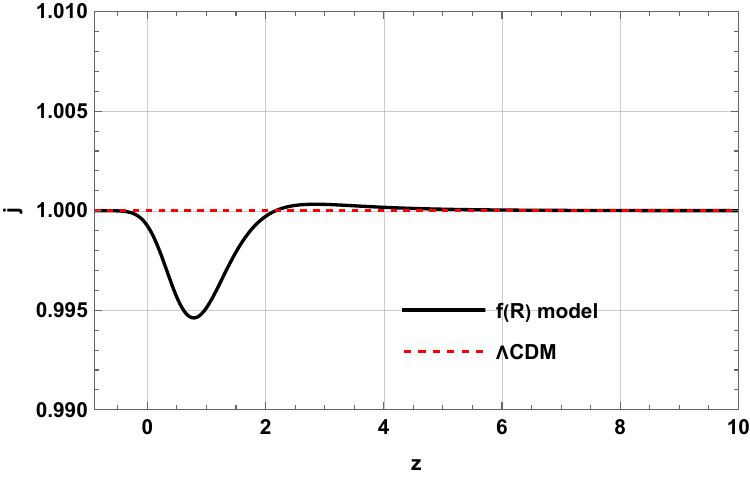}
    \caption{(\emph{left}) Evolution of $q$ vs. $z$  and (\emph{right}) $j$ vs. $z$.}
    \label{fig4}
\end{figure*}														
\begin{figure*}
\centering
    \includegraphics[width=0.47\textwidth]{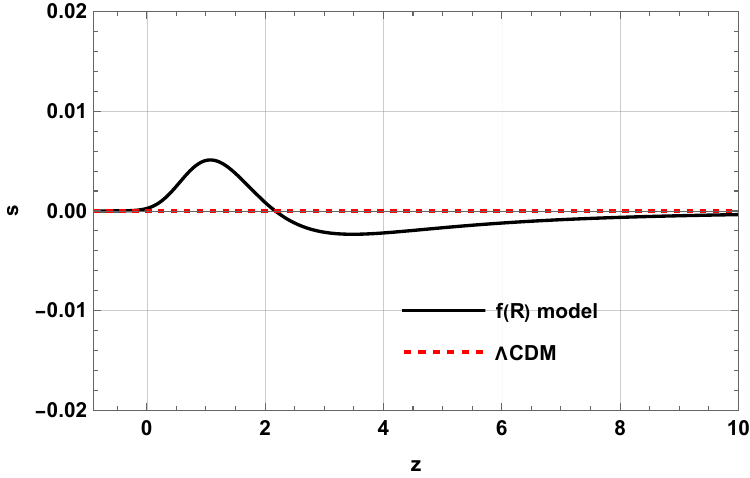}
    \includegraphics[width=0.47\textwidth]{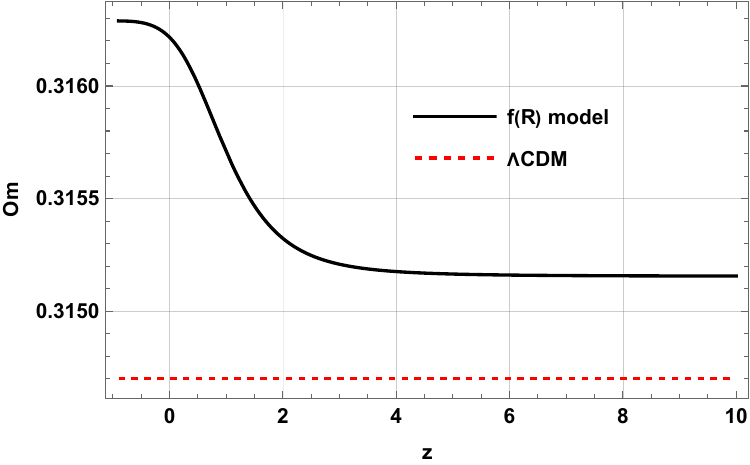}
    \caption{(\emph{left}) Evolution of $s$ vs. $z$  and (\emph{right}) $Om$ vs. $z$.}
    \label{fig5}
\end{figure*}
		
\begin{figure*}
\centering
    \includegraphics[width=0.47\textwidth]{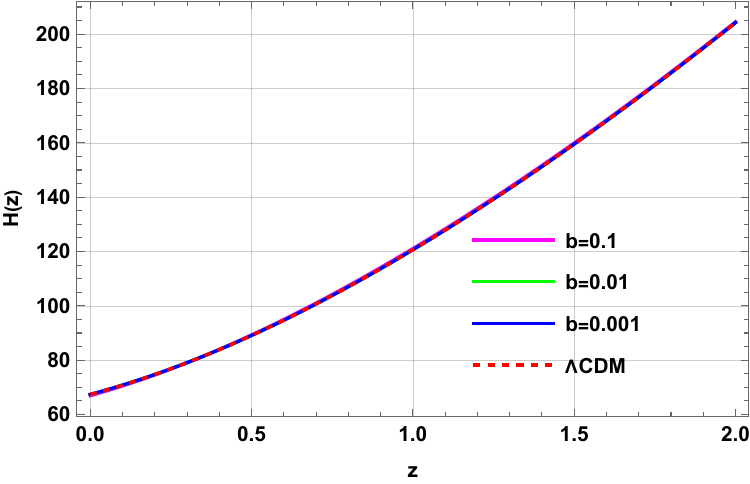}
    \includegraphics[width=0.47\textwidth]{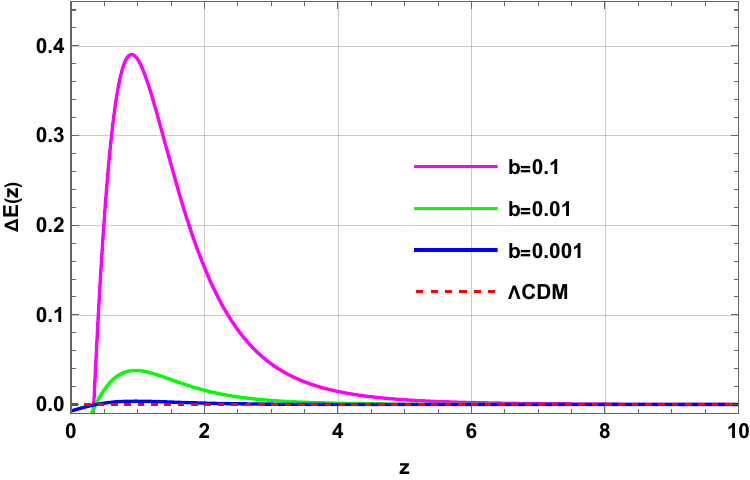}
    \caption{(\emph{left}) Evolution of the Hubble parameter, $H(z)$, (in units of $\rm{km}\,\rm{s}^{-1}\,\rm{Mpc}^{-1}$) using the approximated analytical solution given by Eq. (\ref{eq22}) and the $\Lambda$CDM model as functions of the red shift.
(\emph{right}) Comparison of our model against $\Lambda$CDM using the dimensionless Hubble parameter through Eq. (\ref{eq31}).}
    \label{fig6}
\end{figure*}

\noindent In order to compare our model against the $\Lambda$CDM model, we use the dimensionless Hubble parameter, $E(z)=H(z)/H_0$, and calculating
\begin{equation}\label{eq31}
\Delta E(z)=\left[\frac{E(z)}{E(z)_{\Lambda\rm{CDM}}}-1\right]\times 100,
\end{equation}	
which, by definition, is zero for the $\Lambda$CDM model (red-dashed line in the right panel of Fig. \ref{fig6}). Observing Fig. \ref{fig6}
(left panel), it is evident that certain choices of $b$ result in an evolution of $H(z)$ that closely resembles the $\Lambda$CDM model. To quantify this similarity, the left panel of Fig. \ref{fig6} clearly demonstrates that as $b$ decreases, $H(z)$ progressively converges towards the $\Lambda$CDM model. The largest deviation from $\Lambda$CDM occurs around $z\simeq 0.90$ and for $b=0.1$, being $\Delta E(z\simeq0.90)\simeq 0.39\%$. Additionally, regardless of the value of $b$, the approximated analytical solution for $H(z)$ tends to  the
$\Lambda$CDM model at higher redshifts.

\noindent Table \ref{tab1} provides a summary of the current ($z=0$) values obtained for the analyzed cosmological and statefinder parameters. We also include, for comparison purposes, the predicted values by the $\Lambda$CDM model and the available measurements for some of the parameters.									
\begin{table*}[h]
\centering
\caption{Current values obtained for the cosmological and statefinder parameters utilizing the approximate analytical solution for $H(z)$ provided by Eq. (\ref{eq22}), along with fixed values for the free parameters of the $f(R)$ gravity model.  The values for the $\Lambda$CDM model are obtained using Eq. (\ref{eq18}) in Eqs. (\ref{eq24})-(\ref{eq30}). The units for $H_0$ are $\rm{km}\,\rm{s}^{-1}\,\rm{Mpc}^{-1}$. }
{\begin{tabular}{@{}ccccc@{}} \hline
Parameter & $f(R)$& Planck 2018 or SNe Ia & $\Lambda$CDM\\ \hline
$q_0$               & -0.526 & $-0.53^{+0.17}_{-0.13}$ (SNe Ia) & -0.528\\
$j_0$               & 0.999  & -                & 1\\
$s_0$               & 0.0002  & -                & 0\\
$Om_0$              & 0.316 & -                & 0.3147\\
$w_{\rm{DE}0}$      &-0.998  & $-1.03\pm 0.03$    & -1 \\
$\Omega_{\rm{DE}0}$ & 0.686  & $0.6847\pm 0.0073$ & 0.6853\\ 
$w_{\rm{eff}0}$     &-0.684 & -                  & -0.685\\
$H_0$               & 67.351  & $67.4\pm 0.5$      & 67.4\\ \hline
    \end{tabular}
    \label{tab1}}
\end{table*}																
																
\begin{figure*}
\centering
    \includegraphics[width=0.47\textwidth]{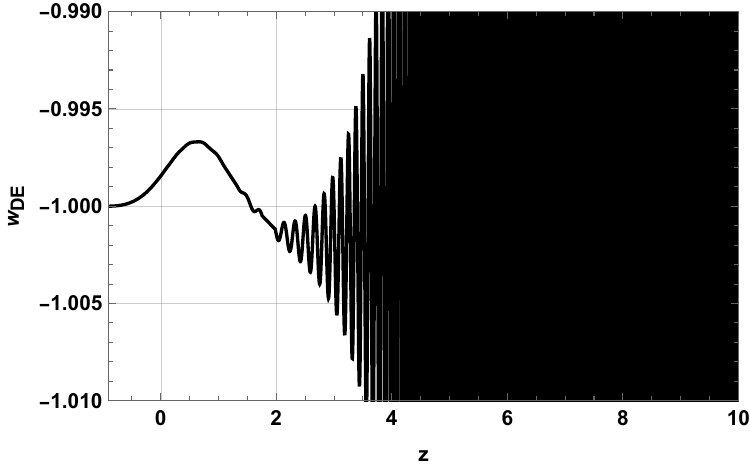}
    \includegraphics[width=0.47\textwidth]{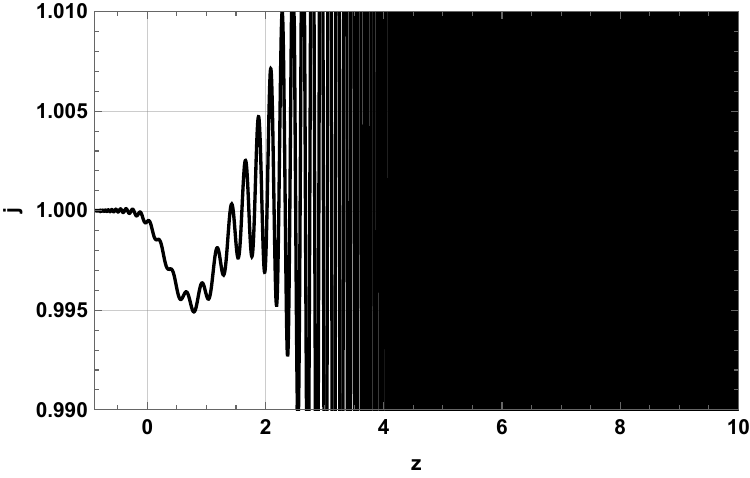}
    \caption{(\emph{left}) Evolution of $w_{\rm{DE}}$ vs. $z$ and (\emph{right}) $j$ vs. $z$. In this case, the numerical solution employed in this study was obtained by utilizing the same initial conditions and model parameter values as those utilized in Ref. \cite{oliveros}, but
taking into account the model given by Eq. (\ref{eq9}),	along with $b=0.01$ and $n=1$.}
    \label{fig7}
\end{figure*}

\noindent It is important to note that the cosmological and statefinder parameters examined earlier do not exhibit significant large frequency oscillations of effective dark energy. Such oscillations would lead to singular and unphysical solutions at high redshifts.  In fact,  if we were to use the numerical solution for this model, the oscillations of effective dark energy would be unavoidable, as depicted in Fig. \ref{fig7}, for instance (see Fig. \ref{fig2} (left panel) and Fig. \ref{fig4} (right panel), for comparison). For simplicity, we have omitted the other plots in this discussion, as they exhibit similar behavior to those presented in Ref. \cite{oliveros}.\\

\noindent Although, as evident in Fig. \ref{fig7}, the presence of oscillations is manifest in our model, these arise when we solve the associated ordinary differential equation numerically for the statefinder quantity $y_{\rm{H}}(z)$, which as has been demonstrated in the literature (see Ref. \cite{bamba}), for larger values of redshift $z$ (matter dominated era) it exhibits oscillations in a similar way as the effective dark energy (since by definition, $y_{\rm{H}}(z)$ is directly related to the effective dark energy). This anomalous behavior is more representative in those cosmological and statefinder parameters which depends on higher derivatives of the Hubble parameter. This problem originates from the stability conditions to be satisfied by $f(R)$ gravity models  and from dark energy oscillations during the matter phase (see Ref. \cite{elizalde1}). In the above context, this behavior is a generic feature of viable $f(R)$ gravity models, like the one proposed here. In our case, we have avoided this anomalous behavior by utilizing the approximate analytical solution for $H(z)$ provided by Eq. (\ref{eq22}), which has not shown large frequency oscillations of effective dark energy, since by construction the solution for $H(z)$ arise from a perturbative approximation to the corresponding $\Lambda$CDM solution for $H(z)$ (see Eq. (\ref{eq15})).  Besides, the expression for $H(z)$ has not oscillatory terms (see Eq. (\ref{eq22})), since the corrective terms to first and second order in $b$ depend explicitly on the $H_{\Lambda}(z)$ and its derivatives (see Eqs. (\ref{eq20})  and (\ref{eq21})), and as is well-known $H_{\Lambda}(z)$ and its derivatives present a good behavior for small and large redshifts, without oscillations.
\subsection{Growth rate of the matter density perturbations}\label{growth}															
\begin{figure*}
\centering
    \includegraphics[width=0.47\textwidth]{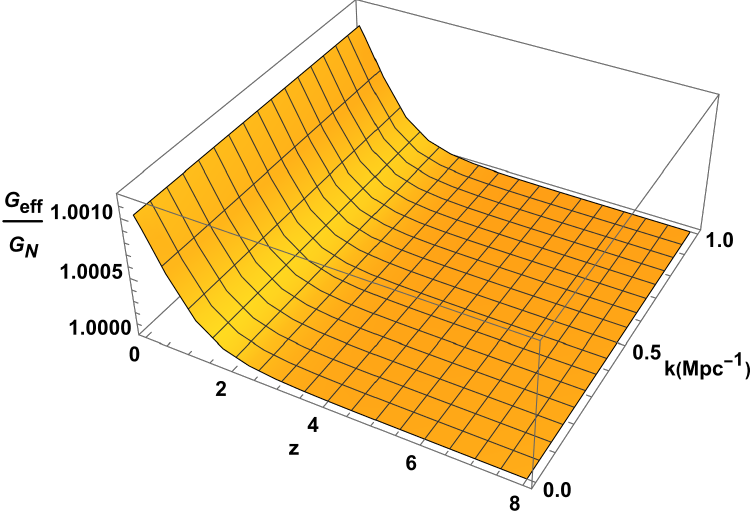}
    \includegraphics[width=0.47\textwidth]{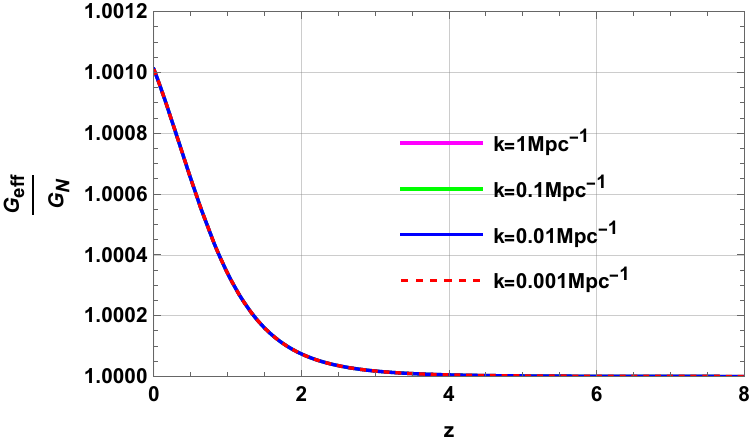}
    \caption{(\emph{left}) Evolution of $G_{\rm{eff}}/G_{\rm{N}}$ as a function of $z$ and the scale dependence on $k$ and (\emph{right})
$G_{\rm{eff}}/G_{\rm{N}}$ vs $z$ using some fixed values of $k$. Note that in both cases we have replaced $a$ by $1/(1+z)$ in
Eq. (\ref{eq34}) }
    \label{fig8}
\end{figure*}

\begin{figure*}
\centering
    \includegraphics[width=0.47\textwidth]{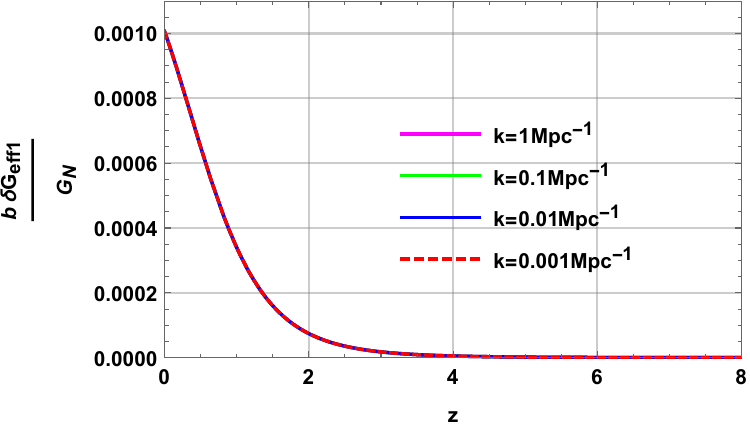}
    \includegraphics[width=0.47\textwidth]{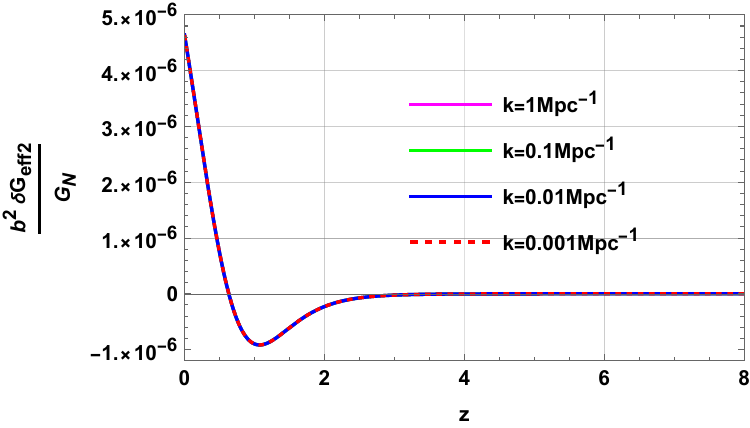}
    \caption{(\emph{left}) Evolution of $b\,\delta G_{\rm{eff1}}/G_{\rm{N}}$ vs. $z$ and (\emph{right})
$b^2\,\delta G_{\rm{eff2}}/G_{\rm{N}}$ vs. $z$, using in both cases some fixed values of $k$.}
    \label{fig9}
\end{figure*}

\begin{figure*}
\centering
    \includegraphics[width=0.47\textwidth]{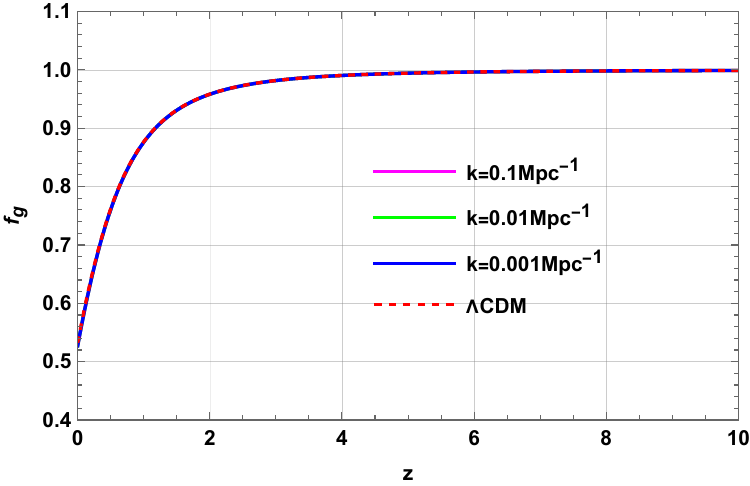}
    \includegraphics[width=0.47\textwidth]{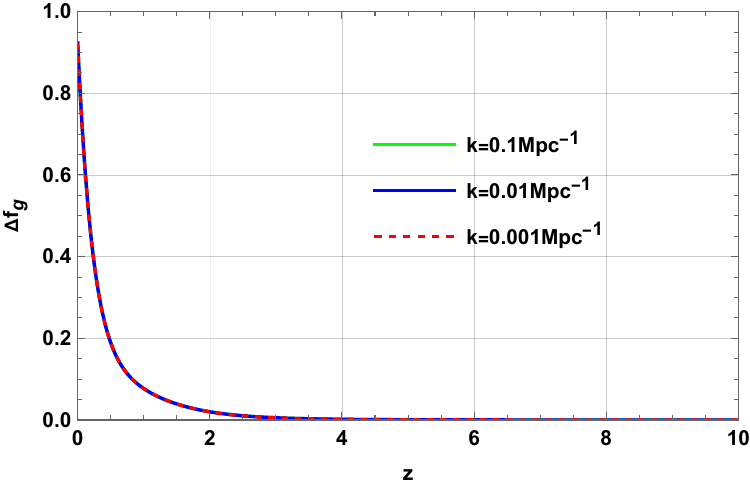}
    \caption{(\emph{left}) Evolution of $f_{\rm{g}}$ vs. $z$ and (\emph{right})
evolution of the relative difference $\Delta f_{\rm{g}}$ vs. $z$, using in both cases some fixed values of $k$ and
the initial condition $f_{\rm{g}}(z_i=50)=1$.}
    \label{fig10}
\end{figure*}

\begin{figure*}
\centering
    \includegraphics[width=0.47\textwidth]{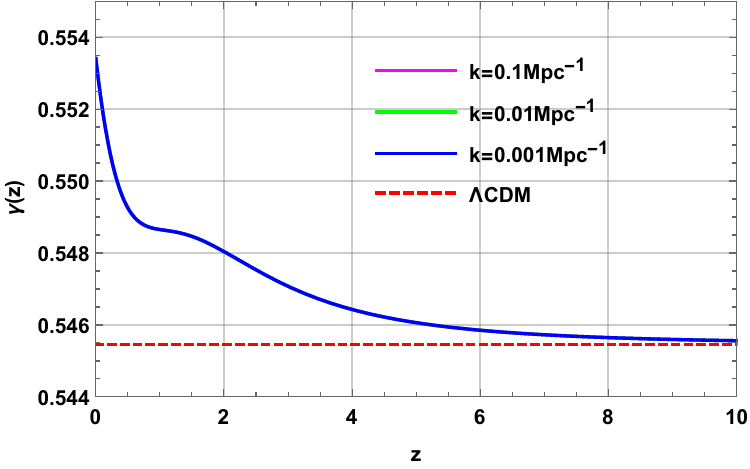}
    \includegraphics[width=0.47\textwidth]{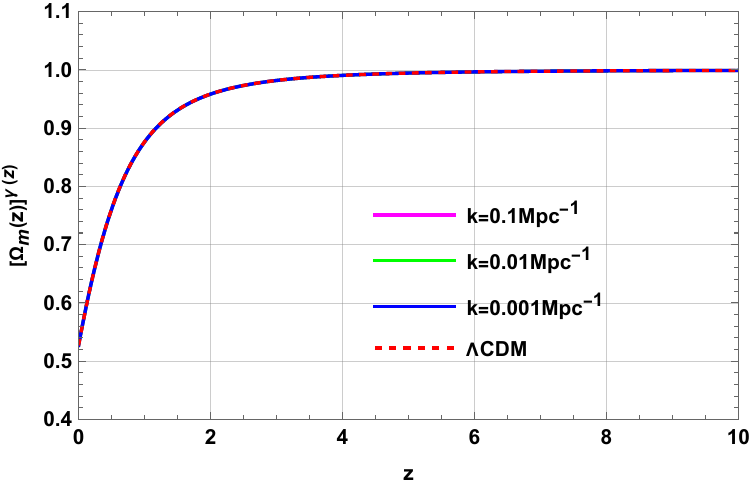}
    \caption{(\emph{left}) Evolution of the growth index $\gamma$ as a function of redshift $z$, using
the initial condition $\gamma(z_i=50)=6/11$. (\emph{right})
$[\Omega_m(z)]^{\gamma(z)}$ vs. $z$, using in both cases some fixed values of $k$.}
    \label{fig11}
\end{figure*}

\noindent To complement the ongoing study, we proceed in this section to analyze the evolution of matter density perturbations for the
$f(R)$ gravity model introduced in Eq. (\ref{eq9}). The equation governing the evolution of matter density perturbations for the $f(R)$ gravity  has been derived previously in the literature, considering the subhorizon approximation ($k^2/a^2\gg H^2$) \cite{tsujikawa1, tsujikawa2, tsujikawa3}.
The gauge-invariant matter density perturbation defined by
\begin{equation}\label{eq32}                                                                
\delta_{\rm{m}}\equiv\frac{\delta\rho_{\rm{m}}}{\rho_{\rm{m}}},
\end{equation}
satisfies the following equation:
\begin{equation}\label{eq33}
\ddot{\delta}_{\rm{m}}+2H\dot{\delta}_{\rm{m}}-4\pi G_{\rm{eff}}(a,k)\rho_{\rm{m}}\delta_{\rm{m}}=0,
\end{equation}
with $k$ being the comoving wavenumber and $G_{\rm{eff}}(a,k)$ being the effective gravitational ``constant'' given by
\begin{equation}\label{eq34}
G_{\rm{eff}}(a,k)=\frac{G_{\rm{N}}}{f_R}\left[1+\frac{(k^2/a^2)(f_{RR}/f_R)}{1+3(k^2/a^2)(f_{RR}/f_R)}\right],
\end{equation}                                                       
in the context of the general theory of relativity, it is important to note that the evolution of matter density perturbations is independent of the comoving wavenumber $k$. However, in $f(R)$ gravity, this dependence arises in the effective gravitational constant, as evident in Eq. (\ref{eq34}). Consequently, the effective gravitational constant generally exhibits scale dependence. 
In Fig. \ref{fig8} (left panel), we illustrate the evolution of normalized effective gravitational constant as a function of the redshift
$z$ and the scale dependence on the comoving wavenumber $k$ for our model. In the right panel, we observe its evolution while considering specific fixed values of $k$. In both figures, it is evident that for approximately $z>4$, the normalized effective gravitational constant tends to 1. This means that for high redshifts, $G_{\rm{eff}}(z,k)$ approaches $G_{\rm{N}}$, which aligns with the prediction of $\Lambda$CDM. Additionally, for $0\leq z<4$, there is a slight variation in $G_{\rm{eff}}/G_{\rm{N}}$ compared to the $\Lambda$CDM prediction.
Furthermore, these results demonstrate a high level of insensitivity to changes in the comoving wavenumber $k$. However, it should be noted that this does not imply that $G_{\rm{eff}}(z,k)$ is completely independent of $k$. In order to illustrate the above, we expand Eq.
(\ref{eq34}) in Taylor series up to second order in $b$:
\begin{equation}\label{eq35}
G_{\rm{eff}}(z,k)\approx G_{\rm{N}}+b\,\delta G_{\rm{eff1}}(z,k)+b^2\,\delta G_{\rm{eff2}}(z,k),
\end{equation} 
where $\delta G_{\rm{eff1}}(z,k)$ is given by
\begin{equation}\label{eq36}
\delta G_{\rm{eff1}}(z,k)=\frac{2G_{\rm{N}}(\Omega_{m0}-1)^2[2k^2(1+z)^2+3H_0^2(4+(-3+z(3+z(3+z)))\Omega_{m0}]}
{3H_0^2[4+(-3+z(3+z(3+z)))\Omega_{m0}]^3},
\end{equation} 
and $\delta G_{\rm{eff2}}(z,k)$ is
\begin{equation}\label{eq37}
\begin{aligned}
\delta G_{\rm{eff2}}(z,k)=&2G_{\rm{N}}(\Omega_{m0}-1)^3\bigg[-\frac{1}{H_0^4}8k^4(1+z)^4(\Omega_{m0}-1)(4+(-3+z(3+z(3+z)))\Omega_{m0})^2\\
&+3(4+(-3+z(3+z(3+z)))\Omega_{m0})(-2+(3+z(3+z(3+z)))\Omega_{m0})(320\\
&+\Omega_{m0}(-592+368z(3+z(3+z))+8(51+23z(3+z(3+z))(-2+z(3\\
&+z(3+z))))\Omega_{m0}+(-135+z(3+z(3+z))(3+z(3+z(3+z))(-181+z\\
&\times(3+z(3+z)))))\Omega_{m0}^2))+\frac{1}{H_0^2}k^2(1+z)^2(-2048+\Omega_{m0}(6848-1344z(3+z\\
&\times(3+z)-24(345+z(3+z)(3+z))(-166+z(3+z(3+z))))\Omega_{m0}+16(297\\
&+z(3+z))(-144++z(3+z(3+z))(105+34z(3+z(3+z)))))\Omega_{m0}^2+(-1269\\
&+z(3+z(3+z))(-324+z(3+z(3+z))(-1638+z(3+z(3+z))(-532\\
&+3z(3+z(3+z))))))\Omega_{m0}^3))\bigg]/[3(4+(-3+z(3+z)))\Omega_{m0}^8].
\end{aligned}
\end{equation}               
In this way, it becomes clear that by employing our model in Eq. (\ref{eq34}), we obtain an explicit dependence on the comoving
wavenumber, $k$. Figure \ref{fig9} illustrates the evolution of the second and third terms in Eq. (\ref{eq35}), which are given by
Eqs. (\ref{eq36}) and (\ref{eq37}). From these figures, we can observe that $b\,\delta G_{\rm{eff1}}/G_{\rm{N}}\ll 1$ and 
$b^2\,\delta G_{\rm{eff2}}/G_{\rm{N}}\ll 1$. Therefore, the second and third terms in Eq. (\ref{eq35}) represent very small corrections to the Newton gravitational constant, $G_{\rm{N}}$. These results better demonstrate the behavior observed in Fig. \ref{fig8} (right panel). 
In principle, the other quantities studied here, such as the cosmological and statefinder parameters, could be expressed through a series expansion in the parameter $b$, where the first term would correspond to the prediction of the $\Lambda$CDM model. However, for the sake of simplicity, we have omitted the explicit writing of these expressions. In Ref. \cite{nesseris1} a series expansion in the parameter $b$
for $w_{\rm{DE}}$ is present, but for the HS model.\\
\noindent In the literature it is a common practice to rewrite Eq. (\ref{eq33}) in terms of the growth rate, which is defined by
\begin{equation}\label{eq38}
f_{\rm{g}}\equiv\frac{d\ln{\delta_m}}{d\ln{a}},
\end{equation} 
where in this case ``g'' stands for ``growth'' and allows us to differentiate of the $f(R)$ function. Realizing the replacements $dt=-\frac{1}{H(1+z)}dz$ and $a=\frac{1}{1+z}$ in Eq. (\ref{eq33}) and after some simple algebraic  manipulations, it take the equivalent form:
\begin{equation}\label{eq39}
\frac{df_{\rm{g}}(z)}{dz}+\left(\frac{1+z}{2E^2(z)}\frac{dE^2(z)}{dz}-2-f_{\rm{g}}(z)\right)\frac{f_{\rm{g}}(z)}{1+z}
+\frac{3\Omega_{m0}(1+z)^2}{2E^2(z)}\frac{G_{\rm{eff}}(z,k)}{G_{\rm{N}}}=0,
\end{equation} 
which has been expressed in terms of $E^2(z)$, since this function is known in explicit form in our case. To numerically solve
Eq. (\ref{eq33}), we adopt an initial condition for the growth rate that is consistent with that observed at very high redshifts, matching that of the $\Lambda$CDM model. In Fig. \ref{fig10} (left panel), we illustrate the cosmological evolution of the growth rate $f_{\rm{g}}$ as a function of redshift $z$ for various fixed values of the comoving wavenumber $k$. Once again, we observe that its behavior is largely unaffected by changes in $k$, and it closely resembles the predictions of the $\Lambda$CDM model. In the right panel of Fig. \ref{fig10},
we compare the growth rate $f_{\rm{g}}$ obtained using our model with the growth rate associated with the $\Lambda$CDM model, denoted as $f_{{\rm{g}}\Lambda}$. The growth rate in the $\Lambda$CDM model, $f_{{\rm{g}}\Lambda}$, is derived from Eq. (\ref{eq39}) by substituting $E^2(z)$ with $E_{\Lambda}^2(z)$, which is given by Eq. (\ref{eq18}). Additionally, we consider $G_{\rm{eff}}(z,k) = G_{\rm{N}}$. This comparison it is 
carried out by using
\begin{equation}\label{eq40}
\Delta f_{\rm{g}}=\left[\frac{|f_{\rm{g}}-f_{{\rm{g}}\Lambda}|}{f_{\rm{g}}}\right]\times 100.
\end{equation} 
From Fig. \ref{fig10} (right panel)  is evident that the growth rate $f_{\rm{g}}$ differs slightly from that associated to the $\Lambda$CDM 
model, $f_{{\rm{g}}\Lambda}$, only by $0<z<4$. The maximum value reached for $\Delta f_{\rm{g}}$ is at $z=0$, $\Delta f_{\rm{g}}(z=0)\approx 0.9\%$.\\
There is another way to represent the growth rate $f_{\rm{g}}$ that has been widely used in the literature. It takes on the following form:
\begin{equation}\label{eq41}
f_{\rm{g}}(z)=[\Omega_m(z)]^{\gamma(z)},
\end{equation} 
where $\gamma$ is known as the growth index. In our case $\Omega_m(z)$ is given by
\begin{equation}\label{eq42}
\Omega_m(z)=\frac{\Omega_{m0}(1+z)^3}{E^2(z)}.
\end{equation} 
Since in this work $E^2(z)$ is known, then we can replace Eq. (\ref{eq41}) along with Eq. (\ref{eq42}) in Eq. (\ref{eq39}), and this way we obtain an ordinary differential equation (ODE) for  the growth index, $\gamma(z)$. To solve numerically this equation, we consider again that for very high redshifts, the growth index  matches that of the $\Lambda$CDM model (i.e. the growth index for the $\Lambda$CDM model is $\gamma\approx 6/11$). In Fig. \ref{fig11} (left panel), we have plotted the growth index $\gamma$ as a function of redshift $z$. It is evident that for the model being studied, the growth index is not constant. However, similar to the other quantities analyzed previously, $\gamma(z)$ is practically insensitive to changes in $k$.
In the right panel of Fig. \ref{fig11}, we illustrate the cosmological evolution of the growth rate utilizing Eq. (\ref{eq41}), while considering the previously obtained growth index, $\gamma(z)$. It is worth noting that its behavior closely resembles what was observed in the left panel of Fig. \ref{fig10}.\\
\noindent  In Ref. \cite{bamba}, the authors explored various parameterizations for the growth index $\gamma$ within a cosmological framework where the effective dark energy originates from two specific $f(R)$ gravity models. They considered specific ansatz for the growth index $\gamma$, namely $\gamma=\gamma_0$, $\gamma=\gamma_0+\gamma_1z$, and $\gamma=\gamma_0+\gamma_1\frac{z}{1+z}$. They fitted Eq. (\ref{eq41}) to the solution of Eq. (\ref{eq39}) for different values of the comoving wavenumber $k$ in the two studied $f(R)$ models. However, in our current work, we have not addressed this aspect, as these parameterizations for the growth index will be considered in a forthcoming study
(currently in progress). In that future work, we aim to determine the coefficients of the aforementioned ansatz and the parameters of the model by utilizing a set of observational data to achieve the best possible fit.
\section{Conclusions}\label{conclus} 
\noindent In the recent  years the $f(R)$ gravity it has positioned as an excellent alternative to explain diverse phenomena in the
cosmological and astrophysical context. In this sense, in this work we have proposed a reparameterization of a specific viable $f(R)$ gravity model to represent it as a perturbation from the $\Lambda$CDM model. The $f(R)$ gravity model under consideration includes two parameters, $b$ and $n$, and it is demonstrated that the closeness of this model to $\Lambda$CDM depends on these parameters, allowing for arbitrary proximity (see Eq. (\ref{eq9})). Furthermore, it is shown that the HS model is a limiting case of this reparameterized model (see Eq. (\ref{eq14})). Following the existing literature, we also derive an analytical approximation for the expansion rate $H(z)$ (see Eq. (\ref{eq22})), which shows an excellent agreement between this analytical approximation and the numerical solution over a wide range of redshifts for realistic values of the deviation parameter $b$ (see Fig. \ref{fig1}).
By appropriately selecting values for the model parameters, we have ploted the cosmological parameters $w_{\rm{DE}}$, $w_{\rm{eff}}$, $\Omega_{\rm{DE}}$, and $H(z)$, as well as the statefinder quantities $q$, $j$, $s$, and $Om(z)$ (see Figs. \ref{fig2}, \ref{fig3}, \ref{fig4},
\ref{fig5}, \ref{fig6}). We find that their present values (at $z=0$) are consistent with the observations from Planck 2018 and the values predicted by the $\Lambda$CDM model (see Table \ref{tab1}). It is important to note that the examined cosmological and statefinder parameters do not exhibit significant oscillations of effective dark energy, which could lead to singular and unphysical solutions at high redshifts (see Fig. \ref{fig7}). This anomalous behavior has been avoided by utilizing the approximate analytical solution for $H(z)$.
Additionally, we have performed a detailed analysis of the evolution of matter density perturbations within the introduced $f(R)$ gravity model, and we have observed that its behavior is largely unaffected by changes in $k$, and it closely resembles the predictions of the
$\Lambda$CDM model (see Figs. \ref{fig8}, \ref{fig9}, \ref{fig10}, \ref{fig11}). The results demonstrate that this viable $f(R)$ gravity model is practically indistinguishable from the $\Lambda$CDM model at the background level.\\

\noindent \textbf{Acknowledgements}\\
This work was supported by  Patrimonio Aut\'onomo-Fondo Nacional de Financiamiento para la Ciencia, la Tecnolog\'ia y la Innovaci\'on Francisco Jos\'e de Caldas (MINCIENCIAS-COLOMBIA) Grant No. 110685269447 RC-80740-465-2020, projects 69723 and 69553.

\end{document}